\documentclass[aps,preprint,onecolumn]{revtex4}%
\usepackage{amssymb}
\usepackage{amsmath}
\usepackage{amsfonts}
\usepackage{graphicx}%
\providecommand{\U}[1]{\protect\rule{.1in}{.1in}}

\begin{document}
\title{Quantum entanglement and quantum phase transition in the \textit{XY} model
with staggered Dzyaloshinskii-Moriya interaction}
\author{Fu-Wu Ma}
\author{Xiang-Mu Kong}
\thanks{Corresponding author}
\email{kongxm@mail.qfnu.edu.cn}
\date{\today}
\affiliation{Shandong Provincial Key Laboratory of Laser Polarization and Information
Technology, Department of Physics, Qufu Normal University, Qufu 273165, China }

\begin{abstract}
We study the quantum entanglement and quantum phase transition (QPT) of the
anisotropic spin-$1/2$ \textit{XY} model with staggered Dzyaloshinskii-Moriya
(DM) interaction by means of quantum renormalization group method. The scaling
of coupling constants and the critical points of the system are obtained. It
is found that when the number of renormalization group iterations tends to
infinity, the system exhibit a QPT between the spin-fluid and N\'{e}el phases
which corresponds with two saturated values of the concurrence for a given
value of the strength of DM interaction. The DM interaction can enhance the
entanglement and influence the QPT of the system. To gain further insight, the
first derivative of the entanglement exhibit a nonanalytic behavior at the
critical point and it directly associates with the divergence of the
correlation length. This shows that the correlation length exponent is closely
related to the critical exponent, i.e., the scaling behaviors of the system.

\end{abstract}
\maketitle

\section{INTRODUCTION}

In the quantum systems, entanglement is a pure quantum correlation, which is
the fundamental difference between quantum and classical physics \cite{1}. In
recent years, quantum\ entanglement has attracted much attention in quantum
information theory because of its importance in developing the idea of quantum
computers and other quantum information devices \cite{2,3}. It has also been
realized as a crucial resource to process and send information in different
ways, such as quantum teleportation, quantum cryptography, and algorithms for
quantum computations \cite{41,42,43}. In the condensed-matter physics, it is
very significant to discuss the relation between entanglement and quantum
phase transition (QPT) which has been attracted many researchers to
inverstigate \cite{4,5,51}.

For investigating the properties of many-body systems, the
renormalization-group (RG) method is applied. In the past several decades,
much effort had been investigated in many spin systems using this method. Real
space renormalization group method was applied to discuss the critical points
and phase diagrams of Heisenberg and Blume-Capel models
\cite{RSRG1,R.C.Brower,stella,BC}. These properties of some models were also
discussed by Monte Carlo RG \cite{MC,MC1}. The density-matrix RG method which
is a powerful numerical method is used to study the ground and low-lying
states properties of low-dimensional lattice models. It has been applied
successfully to lots of strongly correlated systems in 1D as well as 2D
systems \cite{Stevenr White,DMRG1,DMRG2,DMRG3}. Recently, the pairwise
entanglement of the system is studied by the quantum renormalization-group
(QGR) method which plays an important role in QPT \cite{QRG1,QRG2}. The
spin-$1/2$ Ising and\ Heisenberg models were investigated by the same method
and it is found that the systems exist QPT and nonanalytic behavior, such as
the discontinuity in the quantum critical points \cite{I1,I2,I3,I4}. For
getting the accurate results, the \textit{XXZ} model with
next-nearest-neighbor interactions are investigated \cite{NNN1,NNN2}. It is
found that the tri-critical point and the phase diagram of the sysytem can be obtained.

Some spin models can be supplemented with a magnetic term which is called
Dzyaloshinskii-Moriya (DM) interaction arising from the spin-orbit coupling.
The DM interaction, which was first proposed by Dzyaloshinskii and Moriya
about half century ago\cite{D,M}, can influence the phase transition and the
critical properties of some systems. The relevance of antisymmetric
superexchange interaction which describes quantum antiferromagnetic system was
introduced by Dzyaloshinskii. Moriya found that such interaction arises
naturally in the perturbation theory in magnetic systems with low symmetry.
The form of DM interaction for two spins $\overrightarrow{S_{i}}$ and
$\overrightarrow{S_{j}}$ is $\overrightarrow{D}\cdot\left(  \overrightarrow
{S_{i}}\times\overrightarrow{S_{j}}\right)  $. Ising and \textit{XXZ} models
with DM interaction were studied in Ref. \cite{I1,I2}. The results are that
the critical points of the systems are obtained and divided the systems into
two phases, i.e., spin-fluid and N\'{e}el phases. At the critical point, the
nonanalytic behavior of the first derivative of the entanglement and the
scaling behavior of the systems are also gotten.

The quantum entanglement and QPT of the spin-$1/2$ \textit{XY} model with
staggered DM interaction are discussed by using QRG method. We find that the
stable and unstable fixed points of the system can be obtained and the phase
transition point changes as the DM\ interaction increases.. The concurrence is
calculated which is influenced by the anisotropy parameter and DM interaction.
The concurrence trends two fixed values which associate with the phases of the
system as the number of RG iteration increases. Furthermore, the first
derivative of the concurrence shows nonanalytic behavior at the critical point
which has relation with the correlation length. This paper is organized as
follows. In Sec. \ref{di er}, we apply QRG method to investigate the model and
obtain the fixed points. The concurrence is introduced to measure the
entanglement and we analysis it in order to get more insights about the
critical features of the model in Sec. \ref{di san}. We summarize in Sec.
\ref{di si}.

\section{QUANTUM RENORMALIZATION GROUP OF THE MODEL \label{di er}}

The mode elimination or the thinning of the degrees of freedom followed by an
iteration, which reduces the number of lattices step by step until reaching a
more tractable circumstance, is the main idea of RG method. RG is a proper
method to give the universal behavior at long wavelengths,\ it includes many
methods, such as decimation, bond-moving and cumulant expansion. In this
paper, the Kadanoff's block approach is implemented where we have consider
three sites as a block (marking as 1-2-3). Generally speaking, this method
includes three steps. Firstly, the system is divided into blocks and the
Hamiltonian of each block can be exactly diagonalized and solved. Then, the
projection operator is builded by the lower eigenvectors. And finally, the
full Hamiltonian is projected onto these eigenvectors to obtain the effective
Hamiltonian which acts on the renormalized subspace, i.e., the RG equations
\cite{subspace1,subspace2}.

The Hamiltonian of 1D anisotropic \textit{XY} model with staggered DM
interaction on a periodic chain of $N$ sites can be written as%
\begin{equation}
H_{0}=\frac{J}{4}\sum_{i=1}^{N}\left[  \left(  1+\gamma\right)  \sigma_{i}%
^{x}\sigma_{i+1}^{x}+\left(  1-\gamma\right)  \sigma_{i}^{y}\sigma_{i+1}%
^{y}+\left(  -1\right)  ^{i}D\left(  \sigma_{i}^{x}\sigma_{i+1}^{y}-\sigma
_{i}^{y}\sigma_{i+1}^{x}\right)  \right]  , \label{1}%
\end{equation}
where $J$ is the nearest exchange coupling constant, $\gamma$ is the
anisotropy parameter, $D$ is the strength of DM interaction in the direction
of $z$, and $\sigma_{i}^{\alpha}$ $\left(  \alpha=x,y\right)  $ are Pauli
operators of the $i$th site. For the $H_{0}$, when $\gamma=D=0$, the model
becomes isotropic \textit{XX} model; when $\gamma=1$ and $D=0$, it turns into
the Ising model which was exactly solved in Ref. \cite{Ising jingjie}.

The initial Hamiltonian $H_{0}$ acts on\ the effective Hilbert space and then
the effective Hamiltonian $H^{\text{eff}}$ can be gotten. The essential
criterion for RG is that $H^{^{\text{eff}}}$ have similar structure to $H_{0}%
$, but we can not get this, i.e., the signs of $\sigma_{i}^{y}\sigma_{i+1}%
^{y}$ and $\sigma_{i}^{y}\sigma_{i+1}^{x}$ terms are changed. To avoid this
and produce a self-similar Hamiltonian, we implement a $\pi$ rotation around
the $x$ axis for all even sites and leave all odd sites unchanged \cite{I4}.
Therefore, the transformed Hamiltonian is obtained as follows,%
\begin{equation}
H=\frac{J}{4}\sum_{i=1}^{N}\left[  \left(  1+\gamma\right)  \sigma_{i}%
^{x}\sigma_{i+1}^{x}-\left(  1-\gamma\right)  \sigma_{i}^{y}\sigma_{i+1}%
^{y}+D\left(  \sigma_{i}^{x}\sigma_{i+1}^{y}+\sigma_{i}^{y}\sigma_{i+1}%
^{x}\right)  \right]  . \label{2}%
\end{equation}

For the Kadanoff's block approach, $H$ can be written as%
\begin{equation}
H=H^{B}+H^{BB}, \label{er}%
\end{equation}
where $H^{B}$ is the block Hamiltonian and $H^{BB}$ is the interblock
Hamiltonian. The explicit forms of $H^{B}$ and $H^{BB}$ are%
\begin{equation}
H^{B}=\sum_{l=1}^{N/3}h_{l}^{B}, \label{3}%
\end{equation}%
\begin{equation}
H^{BB}=\frac{J}{4}\sum_{l=1}^{N/3}\left[  \left(  1+\gamma\right)
\sigma_{l,3}^{x}\sigma_{l+1,1}^{x}-\left(  1-\gamma\right)  \sigma_{l,3}%
^{y}\sigma_{l+1,1}^{y}+D\left(  \sigma_{l,3}^{x}\sigma_{l+1,1}^{y}%
+\sigma_{l,3}^{y}\sigma_{l+1,1}^{x}\right)  \right]  , \label{4}%
\end{equation}
where the $l$th block Hamiltonian is%
\begin{align}
h_{l}^{B}  &  =\frac{J}{4}\left[  \left(  1+\gamma\right)  \left(
\sigma_{l,1}^{x}\sigma_{l,2}^{x}+\sigma_{l,2}^{x}\sigma_{l,3}^{x}\right)
-\left(  1-\gamma\right)  \left(  \sigma_{l,1}^{y}\sigma_{l,2}^{y}%
+\sigma_{l,2}^{y}\sigma_{l,3}^{y}\right)  \right. \nonumber\\
&  \ \ \ \ \left.  +D\left(  \sigma_{l,1}^{x}\sigma_{l,2}^{y}+\sigma_{l,1}%
^{y}\sigma_{l,2}^{x}+\sigma_{l,2}^{x}\sigma_{l,3}^{y}+\sigma_{l,2}^{y}%
\sigma_{l,3}^{x}\right)  \right]  . \label{5}%
\end{align}

In terms of matrix product states \cite{matrix}, we can get the eigenvalues
and eigenvectors by exactly solving $h_{l}^{B}$. The ground states which are
doubly-degeneracy are useful to construct the projection operator and
calculate the entanglement in later. In the standard basis $\left\{
\left\vert \uparrow\uparrow\uparrow\right\rangle ,\left\vert \uparrow
\uparrow\downarrow\right\rangle ,\left\vert \uparrow\downarrow\uparrow
\right\rangle ,\left\vert \uparrow\downarrow\downarrow\right\rangle
,\left\vert \downarrow\uparrow\uparrow\right\rangle ,\left\vert \downarrow
\uparrow\downarrow\right\rangle ,\left\vert \downarrow\downarrow
\uparrow\right\rangle ,\left\vert \downarrow\downarrow\downarrow\right\rangle
\right\}  $, and the degenerate ground states are given by%
\begin{equation}
\left\vert \varphi_{0}\right\rangle =\frac{\sqrt{1+D^{2}}}{\sqrt{2}q}\left[
\frac{-q}{\sqrt{2}\left(  Di+1\right)  }\left\vert \uparrow\uparrow
\downarrow\right\rangle +\frac{\gamma}{Di+1}\left\vert \uparrow\downarrow
\uparrow\right\rangle +\frac{-q}{\sqrt{2}\left(  Di+1\right)  }\left\vert
\downarrow\uparrow\uparrow\right\rangle +\left\vert \downarrow\downarrow
\downarrow\right\rangle \right]  , \label{6}%
\end{equation}%
\begin{equation}
\left\vert \varphi_{0}\right\rangle ^{^{\prime}}=\frac{1}{2}\left[
\frac{-\sqrt{2}\left(  1-Di\right)  }{q}\left\vert \uparrow\uparrow
\uparrow\right\rangle +\left\vert \uparrow\downarrow\downarrow\right\rangle
+\frac{-\sqrt{2}\gamma}{q}\left\vert \downarrow\uparrow\downarrow\right\rangle
+\left\vert \downarrow\downarrow\uparrow\right\rangle \right]  , \label{7}%
\end{equation}
where $q=\sqrt{1+D^{2}+\gamma^{2}},$ $\left\vert \uparrow\right\rangle $ and
$\left\vert \downarrow\right\rangle $ are the basis vectors of $\sigma^{z}$ in
itself representation. The energy corresponding to the ground states is%
\[
e_{0}=-q/\sqrt{2}.
\]

We keep the ground states of $h_{l}^{B}$ to define the effective site. The
effective Hamiltonian $H^{\text{eff}}$ and the Hamiltonian $H$ have in common
the low lying spectrum \cite{h h lian xi}. An exactly implementation of this
is given by the following equation,%
\begin{equation}
H^{\text{eff}}=T^{\dagger}HT. \label{9}%
\end{equation}
In the preceding equation, $T=\prod\limits_{l=1}^{N/3}T_{0}^{l}$ is the
projection operator of the system and the specific form of $T_{0}^{l}$ is%
\begin{equation}
T_{0}^{l}=\left\vert \Uparrow\right\rangle _{l}\left\langle \varphi
_{0}\right\vert +\left\vert \Downarrow\right\rangle _{l}\left\langle
\varphi_{0}\right\vert ^{^{\prime}}, \label{shi}%
\end{equation}
where $\left\vert \Uparrow\right\rangle _{l}$ and $\left\vert \Downarrow
\right\rangle _{l}$ are the renamed states of each block in the effective
space, which can be seen as a different spin-1/2 particle. Here, we consider
only the first-order correction in the perturbation theory. By using the Eq.
(\ref{er}), the Eq. (\ref{9}) can also be written as%
\begin{equation}
H^{^{\text{eff}}}=T^{\dag}\left(  H^{B}+H^{BB}\right)  T=T^{\dag}%
H^{B}T+T^{\dag}H^{BB}T, \label{10}%
\end{equation}
The Pauli matrices in $x$ and $y$ directions of renormalization are obtained
as follows,%
\begin{align}
T_{0}^{l\dag}\sigma_{l,j}^{x}T_{0}^{l}  &  =\xi_{j}\sigma_{l}^{^{\prime}%
x}+\zeta_{j}\sigma_{l}^{^{\prime}y},\nonumber\\
T_{0}^{l\dag}\sigma_{l,j}^{y}T_{0}^{l}  &  =\mu_{j}\sigma_{l}^{^{\prime}x}%
+\nu_{j}\sigma_{l}^{^{\prime}y}\text{ \ }\left(  j=1,2,3\right)  , \label{11}%
\end{align}
where%
\begin{align}
\xi_{1}  &  =\xi_{3}=\frac{1+D^{2}+\gamma}{\sqrt{2\left(  1+D^{2}\right)  }%
q},\text{ }\zeta_{1}=\zeta_{3}=\frac{-\gamma D}{\sqrt{2\left(  1+D^{2}\right)
}q},\nonumber\\
\xi_{2}  &  =-\frac{1}{2\sqrt{1+D^{2}}}-\frac{\sqrt{1+D^{2}}\gamma}{q^{2}%
},\text{ }\zeta_{2}=\frac{D}{2\sqrt{1+D^{2}}},\nonumber\\
\mu_{1}  &  =\mu_{3}=\frac{\gamma D}{\sqrt{2\left(  1+D^{2}\right)  }q},\text{
}\nu_{1}=\nu_{3}=\frac{\gamma-1-D^{2}}{\sqrt{2\left(  1+D^{2}\right)  }%
q},\nonumber\\
\mu_{2}  &  =-\frac{D}{2\sqrt{1+D^{2}}},\text{ }\nu_{2}=-\frac{1}%
{2\sqrt{1+D^{2}}}+\frac{\sqrt{1+D^{2}}\gamma}{q^{2}}. \label{12}%
\end{align}
We substitute them into Eq. (\ref{10}) and obtain the effective Hamiltonian,%
\begin{equation}
H^{\text{eff}}=\frac{J^{^{\prime}}}{4}\sum_{k=1}^{N/3}\left[  \left(
1+\gamma^{^{\prime}}\right)  \sigma_{k}^{x}\sigma_{k+1}^{x}-\left(
1-\gamma^{^{\prime}}\right)  \sigma_{k}^{y}\sigma_{k+1}^{y}+D^{^{\prime}%
}\left(  \sigma_{k}^{x}\sigma_{k+1}^{y}+\sigma_{k}^{y}\sigma_{k+1}^{x}\right)
\right]  , \label{13}%
\end{equation}
where%
\begin{equation}
J^{^{\prime}}=\frac{1+D^{2}+3\gamma^{2}}{2q^{2}}J,\text{ }\gamma^{^{\prime}%
}=\frac{3\gamma+3D^{2}+\gamma^{3}}{1+D^{2}+3\gamma^{2}},\text{ }D^{^{\prime}%
}=-D. \label{14}%
\end{equation}

Because the antisymmetric is the special property of DM interaction, i.e.,
$\overrightarrow{D}_{i,j}=-\overrightarrow{D}_{j,i}$, the stable and unstable
fixed points can be gotten by solving $\gamma^{^{\prime}}\equiv\gamma$. The
stable fixed points locate at $\gamma=\infty$ and $\gamma=0,$ the unstable
fixed point is $\gamma=\pm\sqrt{1+D^{2}}$ which separates the spin-fluid
phase, $\gamma=0$ and $\gamma=\infty$, from the N\'{e}el phase, $0<|\gamma
|<\sqrt{1+D^{2}}$.

\section{ENTANGLEMENT ANALYSIS \label{di san}}

There are many measures for pairwise entanglement
\cite{REE1,N,REE,Concurrence}. In this section, we would like to calculate the
concurrence of pure state where we consider one of the degeneracy ground
states. The density matrix of a ground state is composed, i.e.,%
\begin{equation}
\rho=\left\vert \varphi_{0}\right\rangle \left\langle \varphi_{0}\right\vert .
\label{15}%
\end{equation}
The results of using $\left\vert \varphi_{0}^{^{\prime}}\right\rangle $ to
construct the density matrix will be same as Eq. (\ref{15}).

There are two cases to define the concurrence for a three-site block. $\left(
i\right)  $ The concurrence between sites $1$ and $3$ is obtained by summing
over the degrees of freedom of the middle site $2$. $\left(  ii\right)  $ We
trace over the site $1$ or $3$ and get the concurrence between the middle site
$2$ and the other one. Without loss of generality, we consider the case
$\left(  i\right)  $. In the standard basis $\left\{  \left\vert
\uparrow\uparrow\right\rangle ,\left\vert \uparrow\downarrow\right\rangle
,\left\vert \downarrow\uparrow\right\rangle ,\left\vert \downarrow
\downarrow\right\rangle \right\}  $, the reduced density matrix $\rho_{13}$
for sites $1$ and $3$ can be gotten in Eq. (\ref{15}),%
\begin{equation}
\rho_{13}=\text{Tr}_{2}\left[  \rho\right]  =\frac{1+D^{2}}{2q^{2}}\left(
\begin{array}
[c]{cccc}%
\frac{\gamma^{2}}{1+D^{2}} & 0 & 0 & \frac{i\gamma}{D+i}\\
0 & \frac{q^{2}}{2\left(  1+D^{2}\right)  } & \frac{q^{2}}{2\left(
1+D^{2}\right)  } & 0\\
0 & \frac{q^{2}}{2\left(  1+D^{2}\right)  } & \frac{q^{2}}{2\left(
1+D^{2}\right)  } & 0\\
\frac{-i\gamma}{D-i} & 0 & 0 & 1
\end{array}
\right)  . \label{16}%
\end{equation}

$C_{13}$ denotes the concurrence between the sites 1 and 3 which is defined as%
\begin{equation}
C_{13}=\text{max}\left\{  \lambda_{1}-\lambda_{2}-\lambda_{3}-\lambda
_{4},\ 0\right\}  , \label{17}%
\end{equation}
where $\lambda_{k}$ $\left(  k=1,2,3,4\right)  $ are the square roots of
eigenvalues of $R=\rho_{13}\tilde{\rho}_{13}$ in descending order.
$\tilde{\rho}_{13}$ is the spin-flipped state \cite{Concurrence} and its
definition is%
\begin{equation}
\tilde{\rho}_{13}=\left(  \sigma_{1}^{y}\otimes\sigma_{3}^{y}\right)
\rho_{13}^{\ast}\left(  \sigma_{1}^{y}\otimes\sigma_{3}^{y}\right)  ,
\label{18}%
\end{equation}
where $\rho_{13}^{\ast}$ is the complex conjugate of $\rho_{13}$. The value of
$C_{13}$ ranges from $0$ to $1$, if $C_{13}=0$ or $1$, the system is in an
unentangled or a maximally entangled state, else it corresponds to a partial
entangled state. The square eigenvalues of $R$ are gotten,%
\begin{equation}
\lambda_{1}=\frac{1}{2},\text{ }\lambda_{2}=\frac{\gamma\sqrt{1+D^{2}}}%
{q},\text{ }\lambda_{3}=\lambda_{4}=0. \label{19}%
\end{equation}
Thus, the concurrence is obtained as follows,%
\begin{equation}
C_{13}=\text{max}\left\{  \lambda_{1}-\lambda_{2},\ 0\right\}  =\frac{1}%
{2}-\frac{\gamma\sqrt{1+D^{2}}}{q}. \label{20}%
\end{equation}

It is easy to see that $C_{13}$ is influenced by $\gamma$ and $D$. For
three-site model, we plot $C_{13}$ versus $\gamma$ for different values of $D$
in Fig. 1. From the figure, it is found that the entanglement is a fixed value
regardless of any value of $D$ at $\gamma=0$ or infinity. That is to say there
are not phase transition for the \textit{XX} model. Moreover, the DM
interaction plays an important role and enhances the entanglement of the
system when $\gamma$ is small comparing to $D$; the effect of anisotropy
parameter for the entanglement is more important than DM interaction when
$\gamma$ is large.

The purpose of QRG is that the full properties of the model enter a few sites
through the renormalizing of coupling constants. The renormalization of the
strength of DM interaction and anisotropic parameter are obtained which are
contribution to the concurrence of the system. For a fixed value of $D=1$, the
graph for $C_{13}$ and $\gamma$ is plotted in Fig. 2. It reveals that as the
number of QRG iterations increases, the concurrence develops three rather
different features which are separated by $\gamma=0$ and $\gamma=\sqrt{2}$.
When the number of RG iterations is very large, i.e., the system is infinity,
the value of $C_{13}$ is zero corresponding to N\'{e}el phase for $\gamma$
ranging from $-\sqrt{2}$ to $\sqrt{2}$ except the point of $0$; at $\gamma=0$,
the system is the spin-fluid phase corresponding to the maximum value of the
concurrence; for $\left\vert \gamma\right\vert >$ $\sqrt{2}$, the concurrence
of the system slowly increases with $\left\vert \gamma\right\vert $ increasing
and finally reaches to the maximum value. The system is in the same phase for
$\gamma=0$ and $\gamma=\infty$. For this, it can be explained from the
Hamiltonian, i.e., the anisotropic \textit{XY} model changes the isotropic
\textit{XX} model for $\gamma=0$ and $\gamma=\infty$.

Further insight, when the system is large enough, the first derivation of the
entanglement shows the nonanalytic behavior at the critical point. Because
$dC_{13}/d\gamma$ is an even function of $\gamma$, we have only plotted
$dC_{13}/d\gamma$ in $\gamma\geq0$ for $D=1$ in Fig. 3. When the QRG iteration
trends infinity, there are a minimum and a maximum values for each plot. From
the diagram, it is found that the singular behavior of the concurrence becomes
more pronounced at the thermodynamic limit. For the inset of diagram, there is
a maximum value for each curved shape which verges to zero as the system
becomes large. These also manifest that there are same properties for
$\gamma=0$ and $\gamma=\infty$. For a more detailed analysis, the positions of
the minimum or the maximum of $dC_{13}/d\gamma$ with the size of system
increasing are given in Fig. 4. It can be seen that it shows a linear
behavior. That is to say the position of the minimum or maximum of
$dC_{13}/d\gamma$ tends toward the critical point $\gamma=\sqrt{1+D^{2}}$. The
critical exponent $\theta$ for this behavior is $dC_{13}/d\gamma
|_{\gamma_{\min}}$ or $dC_{13}/d\gamma|_{\gamma_{\max}}\sim N^{0.98}$. These
results justify that $\theta$ is the reciprocal of the correlation length
exponent $\upsilon$ closing to the critical point, i.e., $\theta=1/\upsilon$.

Next, we investigate how the concurrence versus $D$ changes for a fixed value
of $\gamma$. The concurrence with $D$ changing for different iterations is
depicted in Fig. 5. It is clearly seen that the concurrence becomes large with
$D$ increasing and then until to $0.5$ at last. But the larger the size of
system is, the slower the increasing tendency of the concurrence is. This
implies that the system can not occur QPT by changing the DM interaction. The
first partial derivative of concurrence for $D$ is clearly discussed in Fig.
6. There is not suddenly mutative in the diagram, but the maximum value of
each plot becomes small with the size of the system largening. While we can
also get the linear behavior between the maximum of $dC_{13}/dD$ and the
corresponding size of system which is plotted in Fig. 7. The exponent for the
curve is $dC_{13}/dD|_{D_{\max}}\sim N^{-0.98}$. It falls into the Ising-like
universality class. The position of the maximum of concurrence tends to
infinity as the iteration of QRG increases, in other words, the stable fixed
point $D\rightarrow\infty$ is reached.

\section{CONCLUSIONS \label{di si}}

The relation between the concurrence as a measure of quantum correlations and
QPT was obtained. The anisotropy and DM interaction parameters determined the
phase diagrams of the model.\ If the anisotropy parameter was small, the
concurrence depended on the DM interaction, else the anisotropy parameter
played an important role. As the number of RG iterations increased, the
concurrence developed two different values which separated two phases for a
given $D$, i.e., spin-fluid phase and N\'{e}el phase. For N\'{e}el phase, the
larger the value of $D$ was, the wider the width of value of $\gamma$ was. The
critical behavior was described by the first derivative of the concurrence of
the blocks. The scaling behavior characterizes how the critical point of the
model was touched as the size of system increased. The critical exponent had
relation with the correlation length exponent in the vicinity of the critical
point. This implied that the quantum critical properties of the model were
closely associated with the behavior of the entanglement. The concurrence
increased slowly as the the size of the system became large, but the tendency
of concurrence was unanimous, namely trending a fixed value.

\newpage%
\[
\text{\textbf{Figure Captions}}%
\]

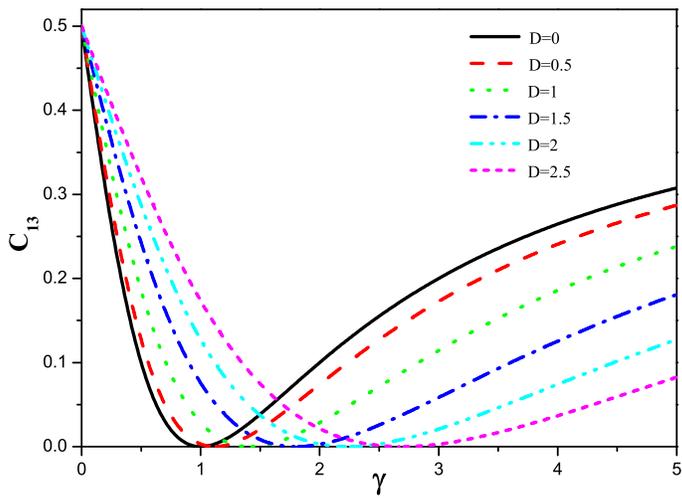
Fig 1. (Color online) Concurrence between the first and third sites of a
three-site model in terms of anisotropy for different values of the strength
of DM interaction.

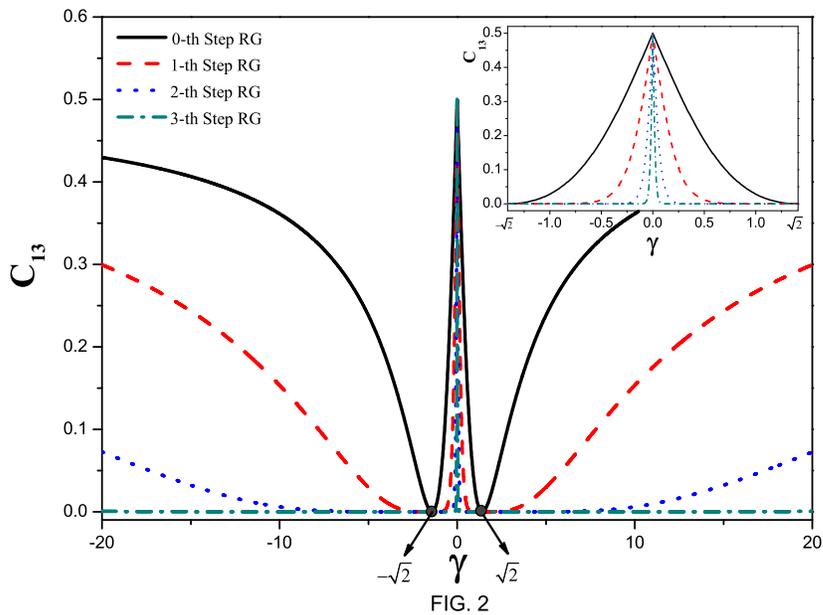
Fig 2. (Color online) Representation of the evolution of the concurrence in
terms of QRG iterations at a fixed value of $D=1$. The inset of diagram of
$\gamma$ ranges from $-\sqrt{2}$ to $\sqrt{2}$.

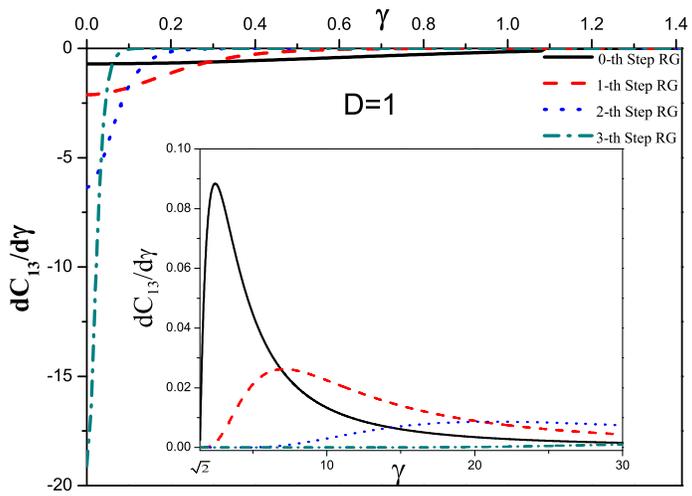
Fig 3. (Color online) First derivative of concurrence and its manifestation
toward diverging as the number of QRG iterations increases (Fig. 2). The value
of the inset of diagram manifestation toward zero.

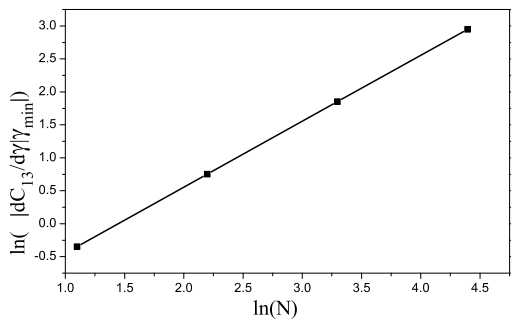
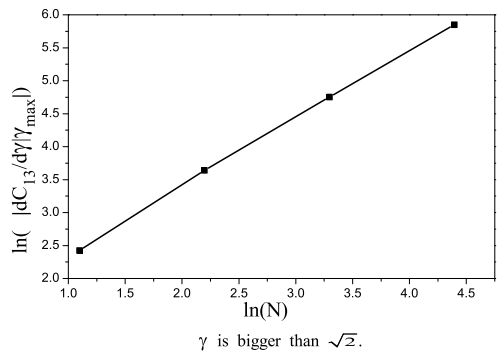
Fig 4. (Color online) The logarithm of the absolute value of minimum or
maximum, ln$(|dC_{13}/d\gamma|\gamma_{m}|)$, versus the logarithm of chain
size, ln$(N)$, which shows a scaling behavior. Each point corresponds to the
minimum or maximum value of a single plot of Fig. 3.

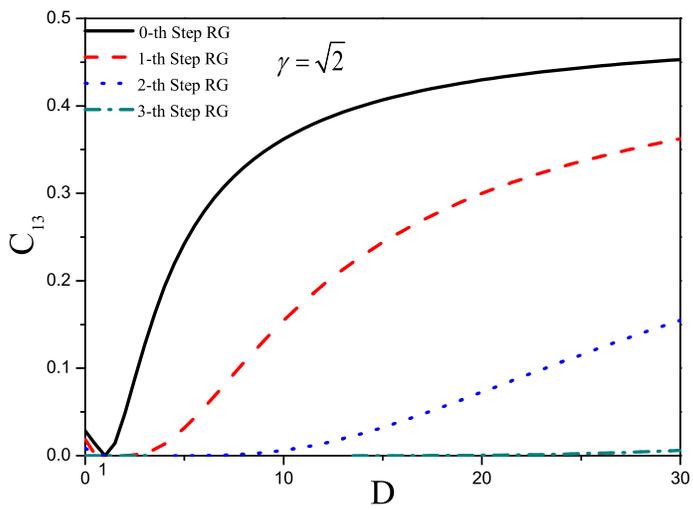
Fig 5. (Color online) Representation of the evolution of the concurrence in
terms of QRG iterations at a fixed value of $\gamma=\sqrt{2}$.

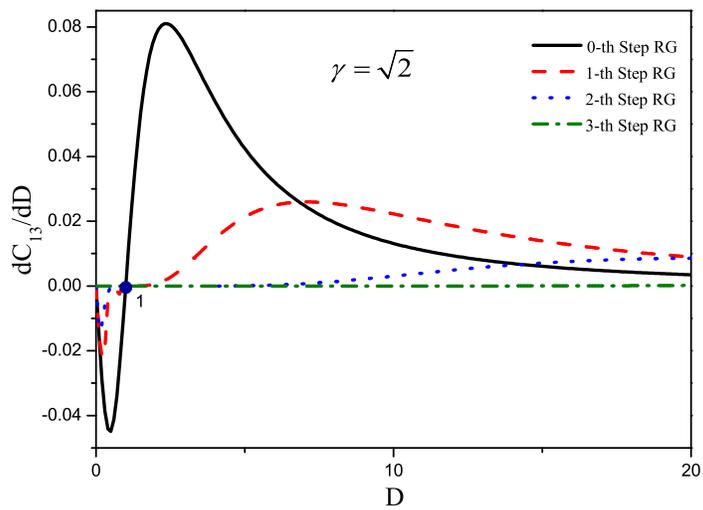
Fig 6. (Color online) First derivative of concurrence and its manifestation
toward zero as the number of QRG iterations increases (Fig. 5).

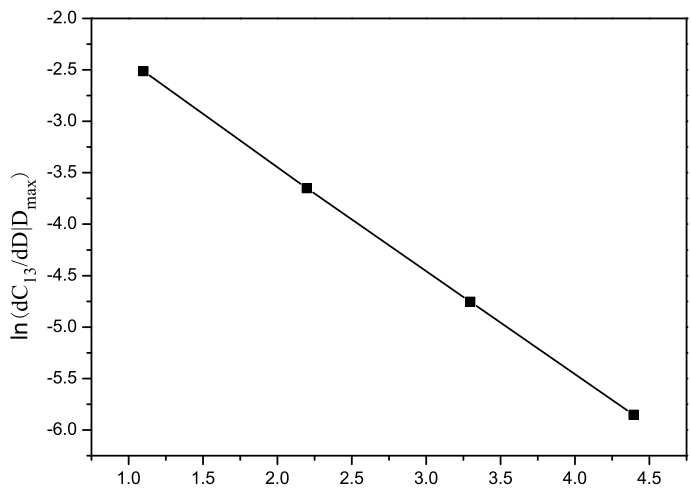
Fig 7. (Color online) The logarithm of the absolute value of maximum,
ln$(|dC_{13}/dD|D_{\max}|)$, versus the logarithm of chain size, ln$(N)$,
which is linear and shows a scaling behavior. Each point corresponds to the
maximum value of a single plot of Fig. 5.

\end{document}